\documentclass[aps,pra,twocolumn,nofootinbib,10pt]{revtex4-1}

\usepackage{natbib}
\usepackage{amsmath,amssymb,bm}
\usepackage{dsfont}
\usepackage{graphics}

\usepackage{color}

\newcommand{\tr}{\text{tr}}

\newcommand{\Dcirc}{{\cal D}^{\circ}}

\begin{document}

\title{Spatiotemporal effects in heralded state preparation}
\author{Filippus S. \surname{Roux}}
\affiliation{University of Kwazulu-Natal, Private Bag X54001, Durban 4000, South Africa}
\email{stefroux@gmail.com}

\begin{abstract}
Heralding, which is often used for preparing quantum optical states, is studied to determine the effects of the spatiotemporal properties of the process. Incorporating all the spatiotemporal degrees of freedom, we follow a Wigner functional approach to consider cases where these states are prepared to have Wigner functionals with negative regions, being suitable resources for quantum information technologies. General expressions are derived for single-photon-subtracted and single-photon-added states. As examples, we consider the photon-subtracted squeezed vacuum state, the photon-added coherent state, and the photon-added thermal state. The Wigner functional approach reveals the importance of the spatiotemporal transformations imposed by the experimental conditions.
\end{abstract}

\maketitle

\section{\label{intro}Introduction}

Quantum information technology involves exotic measurements performed on the outputs from quantum processing systems, applied to exotic quantum states serving as quantum resources. In photonic quantum information systems, such quantum resources often require sophisticated preparation techniques, such as heralded quantum state preparation \cite{dakna0,lvovsky}. The quality of such quantum information systems depends on the quality of the quantum states used as resources.

A heralded quantum optical state is produced conditioned on a suitable detection made on a portion of an input state. The portion is separated off from the an input state by some means and then subjected to a heralding measurement. A successful detection registered by the heralding measurement system signals that the remainder of the state has attained the required form or property. Such heralding processes have been used to prepare Fock states \cite{lvovfock,ohtfock1,ohtfock2}, photon subtracted states \cite{biswas,trepstheorem}, and photon added states \cite{tara,zav3}. Such states are considered to be suitable resources for quantum information process. Their quantum nature is associated with the fact that their Wigner functions have negative regions. Photon subtraction is also used to produce states with Wigner functions that resemble those of Schr{\"o}dinger cat states. In the limit of a small number of photons, a photon subtracted state is equivalent to a Schr{\"o}dinger cat state \cite{dakna,homokat1,homokat2}.

The basic mechanism of heralded state preparation involves only the particle-number degrees of freedom \cite{toolbox}. Although this mechanism provides some insight into the quantum properties of the states that are produced, the other degrees of freedom also play significant roles in such experiments \cite{lvovsky0}. The effects of various experimental parameters can only be revealed through an analysis that incorporates the spatiotemporal degrees of freedom.

Here, the effects of the spatiotemporal degrees of freedom in heralded state preparation is investigated by analyzing the process with a Wigner functional formalism, which combines the spatiotemporal degrees of freedom with the particle-number degrees of freedom \cite{mrowc,stquaderr}. For this purpose, we focus on the applications of heralded state preparation for the experimental implementation of photon subtraction and photon addition. As an example of the former, we consider the heralded preparation of a photon-subtracted squeezed vacuum state \cite{biswas}. Examples of the latter that are considered here are photon-added coherent states \cite{zav1} and photon-added thermal states \cite{zav3}. Inevitably, the incorporation of the spatiotemporal degrees of freedom in the analysis implies a significant increase in the complexity of such an analysis. Fortunately, the complexity can be significantly alleviated with the aid of generating functions \cite{toolbox}. In addition, we introduce some further simplifications that are relevant under general experimental conditions.


\section{\label{herald}Heralding}

First, we provide a general model for the processes whereby states are prepared with the aid of heralding. The idea is that the required state is only produced when a specific measurement is performed successfully. In that case, the measurement {\em heralds} the existence of the required state. The latter is then processed and subsequently measured. The final measurements are done in coincidence with the heralding measurement, ensuring the existance of the heralded state during the final measurements. The requirement for coincidence implies that the heralding measurement and the final measurements combine into a measurement protocol performed on the initial state.

\subsection{\label{mathher}Generic process}

In terms of quantum mechanics, we can express the state that is produced by the heralding process as
\begin{equation}
\hat{\rho}_{\text{her}} = \tr_B\{\hat{P}_B \hat{U}_{\text{bs}} \left(\hat{\rho}_{\text{in}}\otimes\hat{\rho}_{\text{vac}}\right) \hat{U}_{\text{bs}}^{\dag}\} ,
\label{heraldpros}
\end{equation}
where $\hat{P}_B$ is the projection operator for the heralding measurement, $\hat{U}_{\text{bs}}$ is a unitary operator for the (homogeneous or inhomogeneous) beamsplitter, and $\hat{\rho}_{\text{in}}\otimes\hat{\rho}_{\text{vac}}$ is the tensor product of the initial state and the vacuum state, respectively entering the two input ports of the beamsplitter. A partial trace is evaluated over the part of the state that is received from the output port of the beamsplitter on which the projection operator is applied. The difference between a {\em homogeneous beamsplitter} and an {\em inhomogeneous beamsplitter}, is that the former applies the splitting ratio indiscriminately on all photons, and the latter uses the spatiotemporal properties of the photons to determine the splitting ratio. Inhomogeneous beamsplitters come in handy when modeling loss processes introduced by structures such as apertures.

To analize the effects of the physical system, incorporating any physical restrictions associated with the spatiotemporal degrees of freedom of the system, we use Wigner functionals. The effect of the beamsplitter on such Wigner functionals is represented by transformations of the field variables on which they depend.

The heralding measurement is modeled with generalized projection operations expressed in terms of a generating function for the Wigner functionals of such projection operators. It is given by
\begin{equation}
\mathcal{P}(J)
= \mathcal{N}^{\tr\{D\}} \exp\left(-2\mathcal{J}\alpha^*\diamond D\diamond\alpha\right) ,
\label{fotnumgen2}
\end{equation}
where $\alpha$ and $\alpha^*$ represent the field variables. We define
\begin{equation}
\mathcal{N} = \frac{2}{1+J} ~~~ \text{and} ~~~ \mathcal{J} = \frac{1-J}{1+J} ,
\end{equation}
in terms of $J$, the auxilary parameter used to generate the individual Wigner functionals of the projection operators, and use $\diamond$-contractions to represent
\begin{equation}
\alpha^*\diamond D\diamond\alpha
= \int \alpha^*(\mathbf{k}_1)D(\mathbf{k}_1,\mathbf{k}_2)\alpha(\mathbf{k}_2)\frac{d^3 k_1}{(2\pi)^3}\ \frac{d^3 k_2}{(2\pi)^3} ,
\end{equation}
with $D(\mathbf{k}_1,\mathbf{k}_2)$ denoting a {\em detector kernel}.

Heralding is a post-selection process, which is not trace-preserving. As a result, the raw heralded state is not normalized and thus needs to be normalized by dividing it by its trace.

\begin{figure}[ht]
\centerline{\includegraphics{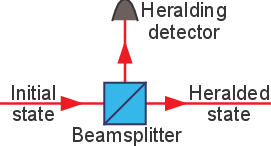}}
\caption{Diagrammatic representation of the experimental setup for heralded photon subtraction.}
\label{fotaf}
\end{figure}

\section{\label{aftrek}Photon subtracted states}

One way to produce a Wigner functional with a negative region is to subtract a photon from a squeezed state; the latter having a Wigner functional that is positive everywhere in the functional phase space. Formally, a photon subtraction is done by applying an annihilation operator to an initial state. In a practical experiment, photon subtraction is performed with the aid of heralding. A small portion of the state is separated off by an {\em unbalanced homogeneous beamsplitter}, having a small reflectivity, and sent to a photon-number resolving detector, as shown in Fig.~\ref{fotaf}. The detection of $n$ photons by this detector then heralds the formation of an $n$-photon subtracted state.

The unitary transformation imposed by the homogeneous beamsplitter on the input state is implement by the transformations of the field variables of the Wigner functional on which it operates. For the unbalanced beamsplitters with a small reflectivity, we represent the {\em amplitude reflectivity} by $\zeta$; the intensity reflectivity is $\zeta^2$. The transformation of the field variables is given by
\begin{align}
\begin{split}
\alpha \rightarrow & \sqrt{1-\zeta^2}\alpha+i\zeta\beta , \\
\beta \rightarrow & \sqrt{1-\zeta^2}\beta+i\zeta\alpha .
\end{split}
\label{wigbstra}
\end{align}
where $\alpha$ and $\beta$ represent the field variables of the respective input ports or the respective output ports. This transformation is applied to the product of the Wigner functionals of the input state and a vacuum state. The Wigner functional of the vacuum state is
\begin{align}
W_{\text{vac}}[\beta] = & \mathcal{N}_0\exp\left(-2\beta^*\diamond\beta\right) ,
\end{align}
where the field variables are $\beta$ and $\beta^*$, and $\mathcal{N}_0$ is a (divergent) normalization constant. After applying the beamsplitter transformation, the combined state becomes
\begin{align}
W_{\text{bs}}[\alpha,\beta] = & \mathcal{N}_0\exp\left[-2\left(\sqrt{1-\zeta^2}\beta^*-i\zeta\alpha^*\right) \right. \nonumber \\
& \left. \diamond\left(\sqrt{1-\zeta^2}\beta+i\zeta\alpha\right)\right] \nonumber \\
& \times W_{\text{in}}\left[\sqrt{1-\zeta^2}\alpha+i\zeta\beta\right] .
\end{align}

Assuming a small reflectivity, we expand the Wigner functional of the state after the beamsplitter as a power series in $\zeta$. Then we use Eq.~(\ref{fotnumgen2}) as a functional of $\beta$ to apply measurements on the $\beta$-degrees of freedom, tracing over $\beta$. It produces a generating function for the Wigner functionals of the heralded state associated with the detection of a certain number of photons. To detect $n$ photons, we need to expand the transformed Wigner functional to $\zeta^{2n}$. Here, we consider single-photon subtraction only, for which we need to make the expansion up the $\zeta^2$. The procedure can be readily generalized for larger numbers of photon subtractions.

All the uneven order terms in the expansion are removed by the trace over $\beta$. The zeroth order term represents the vacuum, which does not contributed to the measurement. So, for single-photon subtraction, we only need to compute the second order term in the expansion. The two derivatives with respect to $\zeta$ produces functional derivatives of the Wigner functional $W_{\text{in}}$. The result for the second order term is
\begin{widetext}
\begin{align}
\tfrac{1}{2}\zeta^2 \left. \partial_{\zeta}^2 W_{\text{bs}}[\alpha,\beta] \right|_{\zeta=0}
= & \tfrac{1}{2}\zeta^2 \left[2\beta^*\diamond\left(\frac{\delta^2 W_{\text{in}}}{\delta\alpha^*\delta\alpha}
+2\alpha\frac{\delta W_{\text{in}}}{\delta\alpha}+2\frac{\delta W_{\text{in}}}{\delta\alpha^*}\alpha^*
+2 W_{\text{in}}\mathbf{1} + 4 \alpha W_{\text{in}}\alpha^* \right)\diamond\beta
-\beta^*\diamond\left(\frac{\delta^2 W_{\text{in}}}{\delta\alpha^*\delta\alpha^*} \right. \right. \nonumber \\
& \left. +4\frac{\delta W_{\text{in}}}{\delta\alpha^*}\alpha
+ 4 \alpha W_{\text{in}}\alpha\right)\diamond\beta^* -\beta\diamond\left(\frac{\delta^2 W_{\text{in}}}{\delta\alpha\delta\alpha}
+4\alpha^*\frac{\delta W_{\text{in}}}{\delta\alpha} + 4 \alpha^* W_{\text{in}}\alpha^*\right)\diamond\beta \nonumber \\
& \left. -\frac{\delta W_{\text{in}}}{\delta\alpha}\diamond\alpha-\alpha^*\diamond\frac{\delta W_{\text{in}}}{\delta\alpha^*}
- 4 \alpha^*\diamond\alpha W_{\text{in}}\right]\mathcal{N}_0 \exp\left(-2\beta^*\diamond\beta\right) ,
\label{tweedord}
\end{align}
\end{widetext}
where $\mathbf{1}$ is the identity, defined such that $\mathbf{1}\diamond\alpha=\alpha$.


For the heralding measurements on the $\beta$-degrees of freedom, we multiply Eq.~(\ref{tweedord}) by Eq.~(\ref{fotnumgen2}), expressed as a functional of $\beta$, and perform a functional integration over $\beta$. The latter produces a superposition of moment integrals of the form
\begin{equation}
M_{m,n} = \int \beta^{*m}\beta^n \mathcal{V}[\beta]\ \Dcirc[\beta] ,
\end{equation}
where
\begin{equation}
\mathcal{V}[\beta] = \mathcal{N}^{\tr\{D\}}
\mathcal{N}_0 \exp\left(-2\beta^*\diamond\beta-2\mathcal{J}\beta^*\diamond D\diamond\beta\right) .
\end{equation}
Such moment integrals can be computed with the aid of a generating function, given by
\begin{align}
\mathcal{M}[\mu^*,\nu]
= & \mathcal{N}^{\tr\{D\}}\mathcal{N}_0 \int \exp\left(-2\beta^*\diamond\beta \right. \nonumber \\
& \left. -2\mathcal{J}\alpha^*\diamond D\diamond\alpha+\beta^*\diamond\nu+\mu^*\diamond\beta\right)\ \Dcirc[\beta] \nonumber \\
= & \exp\left[\tfrac{1}{2}\mu^*\diamond\nu -\tfrac{1}{4}(1-J)\mu^*\diamond D\diamond\nu\right] .
\label{genmomente}
\end{align}
The moments are produced by functional derivatives with respect to the auxiliary field variables $\mu^*$ and $\nu$ producing factors of $\beta$ and $\beta^*$ respectively, after which the auxiliary field variables are set equal to zero. For example
\begin{align}
M_{1,1} = & \int \beta^*(\mathbf{k})\beta(\mathbf{k}') \mathcal{V}[\beta]\ \Dcirc[\beta] \nonumber \\
= & \left. \frac{\delta^2 \mathcal{M}[\mu^*,\nu]}{\delta\nu(\mathbf{k})\delta\mu^*(\mathbf{k}')} \right|_{\mu^*=\nu=0} \nonumber \\
= & \tfrac{1}{2}\mathbf{1}(\mathbf{k}',\mathbf{k}) -\tfrac{1}{4}(1-J)D(\mathbf{k}',\mathbf{k}) .
\label{bbmomente}
\end{align}
Similarly, we obtain $M_{2,0}=M_{0,2}=0$ and $M_{0,0}=1$. With these moments, the generating function of the heralded single-photon subtracted state becomes
\begin{align}
\mathcal{W}_{1\text{ps}}[\alpha]
= & \tfrac{1}{2}\tr\left\{\frac{\delta^2 W_{\text{in}}}{\delta\alpha^*\delta\alpha}\right\}
+\tfrac{1}{2}\frac{\delta W_{\text{in}}}{\delta\alpha}\diamond\alpha
+\tfrac{1}{2}\alpha^*\diamond\frac{\delta W_{\text{in}}}{\delta\alpha^*} \nonumber \\
& +\Omega W_{\text{in}}
-\tfrac{1}{4}(1-J)\left(\tr\left\{D\diamond\frac{\delta^2 W_{\text{in}}}{\delta\alpha^*\delta\alpha}\right\} \right. \nonumber \\
& +2\tr\{D\} W_{\text{in}} +4\alpha^*\diamond D\diamond\alpha W_{\text{in}} \nonumber \\
& \left. +2\frac{\delta W_{\text{in}}}{\delta\alpha}\diamond D\diamond\alpha
+2\alpha^*\diamond D\diamond\frac{\delta W_{\text{in}}}{\delta\alpha^*} \right),
\label{eenfotgen}
\end{align}
where we discarded the factor of $\zeta^2$, and $\Omega=\tr\{\mathbf{1}\}$. Note that the resulting generating function is linear in the auxiliary parameter $J$. It can therefore only produce the single-photon subtracted state. The result obtained from the generating function still needs to be normalized. Assuming that the detector can be modeled by a single-mode kernel $D(\mathbf{k}_1,\mathbf{k}_2)=M(\mathbf{k}_1)M^*(\mathbf{k}_2)$, which implies that $\tr\{D\}=1$, we obtain the single-photon subtracted state as
\begin{align}
W_{1\text{ps}}[\alpha] = & N\left\{ \left. \partial_J \mathcal{W}_{1\text{ps}} \right|_{J=0} \right\} \nonumber \\
= & \mathcal{N} \left(M^*\diamond\frac{\delta^2 W_{\text{in}}}{\delta\alpha^*\delta\alpha}\diamond M +2 W_{\text{in}} \right. \nonumber \\
& +4\alpha^*\diamond MM^*\diamond\alpha W_{\text{in}}
+2\frac{\delta W_{\text{in}}}{\delta\alpha}\diamond MM^*\diamond\alpha \nonumber \\
& \left. +2\alpha^*\diamond MM^*\diamond\frac{\delta W_{\text{in}}}{\delta\alpha^*} \right) ,
\label{eenfotaf}
\end{align}
for an arbitrary initial state $W_{\text{in}}$, where $N\{\cdot\}$ is a normalization process, and $\mathcal{N}$ is the normalization constant.

The normalization can be computed with the generating function that is obtained by integrating Eq.~(\ref{eenfotgen}) over $\alpha$. For this purpose, we note that, since the inital Wigner functional is normalized,
\begin{equation}
\int W_{\text{in}}\ \Dcirc[\alpha] = 1 .
\end{equation}
If the integrant is a total derivative, the result is 0 because the Wigner functional tend to 0 at infinity. Hence,
\begin{equation}
\int \frac{\delta^2 W_{\text{in}}}{\delta\alpha^*\delta\alpha}\ \Dcirc[\alpha] = 0 .
\end{equation}
With the aid of {\em partial functional integration}, we get
\begin{align}
\int \alpha(\mathbf{k})\frac{\delta W_{\text{in}}}{\delta\alpha(\mathbf{k}')}\ \Dcirc[\alpha]
= & \int \alpha^*(\mathbf{k})\frac{\delta W_{\text{in}}}{\delta\alpha^*(\mathbf{k}')}\ \Dcirc[\alpha] \nonumber \\
= & -\delta(\mathbf{k}-\mathbf{k}') .
\end{align}
Using these identities, we obtain
\begin{align}
\mathcal{N}^{-1} = & \left. \partial_J \int \mathcal{W}_{1\text{ps}}\ \Dcirc[\alpha] \right|_{J=0} \nonumber \\
= & \int \alpha\diamond D\diamond\alpha^* W_{\text{in}}\ \Dcirc[\alpha] - \tfrac{1}{2}\tr\{D\} .
\end{align}
We are left with one integral to evaluate, for which we need to know the initial Wigner functional.

Considering different input states, we note that coherent states remain unaffected by photon subtraction processes. Thermal states are affected, but their Wigner functionals remain positive everywhere. To have negative regions, we need squeezed states as input states.

\subsection{Photon subtracted squeezed vacuum state}

The Wigner functional of a squeezed vacuum state is
\begin{align}
W_{\text{sv}} = & \mathcal{N}_0 \exp\left(-2\alpha^*\diamond A\diamond\alpha
-\alpha^*\diamond B\diamond\alpha^* \right. \nonumber \\
& \left. -\alpha\diamond B^*\diamond\alpha\right) ,
\label{squ}
\end{align}
where $A$ and $B$ are a Hermitian kernel and a symmetric kernel, respectively. The purity of the state ensures that
\begin{equation}
A-B\diamond A^{*-1}\diamond B^* = A^{-1} .
\label{idsqut}
\end{equation}
Substituting Eq.~(\ref{squ}) into Eq.~(\ref{eenfotaf}), we obtain a Wigner functional for a single-photon subtracted squeezed vacuum state in the form of a polynomial Gaussian state that becomes negative at the origin.

To compute the normalization factor, we produce a generating functional for the moments in the same way we did in Eq.~(\ref{genmomente}). For the squeezed vacuum state, this generating functional is given by
\begin{align}
\mathcal{W}_{\mathcal{N}}[\mu^*,\nu]
= & \int \mathcal{N}_0 \exp\left(-2\alpha^*\diamond A\diamond\alpha-\alpha^*\diamond B\diamond\alpha^* \right. \nonumber \\
& \left. -\alpha\diamond B^*\diamond\alpha +\alpha^*\diamond\nu+\mu^*\diamond\alpha\right)\ \Dcirc[\alpha] \nonumber \\
= & \exp\left( \tfrac{1}{2}\mu^*\diamond A\diamond\nu
-\tfrac{1}{4}\nu\diamond A^*\diamond B^*\diamond A^{-1}\diamond\nu \right. \nonumber \\
& \left. -\tfrac{1}{4}\mu^*\diamond A^{-1}\diamond B\diamond A^*\diamond\mu^* \right) .
\label{gennormsq}
\end{align}
The inverse normalization constant thus becomes
\begin{align}
\mathcal{N}^{-1} = & \left. \tr\left\{D\diamond\frac{\delta^2 \mathcal{W}_{\mathcal{N}}}{\delta\nu\delta\mu^*}\right\} \right|_{\mu^*=\nu=0}
- \tfrac{1}{2}\tr\{D\} \nonumber \\
 = & \tfrac{1}{2}\tr\left\{D\diamond(A-\mathbf{1})\right\} = \tfrac{1}{2} M^*\diamond E\diamond M ,
\end{align}
for a single-mode detector, where $E=A-\mathbf{1}$.

The normalized heralded state is now given by
\begin{align}
W_{1\text{ps}}^{(\text{sv})}
= & 2\left(\frac{\left|\alpha^*\diamond M_E + M_B^*\diamond\alpha\right|^2}{\eta} - 1 \right) W_{\text{sq}} ,
\label{eenfotsq}
\end{align}
where we defined
\begin{align}
\begin{split}
M_E = & E\diamond M = (A-\mathbf{1})\diamond M , \\
M_B^* = & B^*\diamond M , \\
\eta = & M^*\diamond E\diamond M .
\end{split}
\end{align}
At the origin ($\alpha=0$), we get $W_{1\text{ps}}^{(\text{sv})}[0]=-2$, showing that there is a region where the Wigner functional is negative. For larger numbers of subtracted photons, the number of regions where the Wigner functional is negative increases, but for even numbers of subtracted photons the origin does not lie in a negative region.

Although the functional phase space on which the Wigner functional is defined is infinite dimensional, the $\diamond$-contractions in the polynomial prefactor in Eq.~(\ref{eenfotsq}) reduces it to a four-dimensional subspace of the functional phase space defined by the two transformed modes $M_E$ and $M_B$. When the Wigner functional is measured with a homodyne tomography process \cite{leonhardt,lvovsky0,homod}, the best results are obtained when the mode of the local oscillator lies within this subspace. Knowledge of the detector mode $M$ and the expressions of the $E$ and $B$ kernels \cite{nosemi} make it possible to compute these transformed modes.

\section{\label{optel}Photon added states}

\begin{figure}[ht]
\centerline{\includegraphics{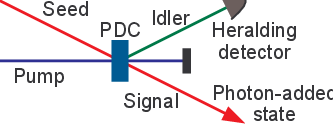}}
\caption{Diagrammatic representation of the experimental setup for heralded photon addition with stimulated parametric down-conversion (PDC).}
\label{fotop}
\end{figure}

The addition of photons to states is more powerful in producing Wigner functionals with negative regions than photon subtraction. Photon addition is formally represented by the application of a creation operator to an initial state. Experimentally, it is done by using the state as the seed in stimulated parametric down-conversion \cite{paramp} and then heralding the photon-added state (in the signal beam) by detecting a certain number of photons in the difference-frequency (idler) beam, as shown in Fig.~\ref{fotop}.

In terms of quantum mechanics, the process of heralded photon addition is represented by
\begin{equation}
\hat{\rho}_{\text{hpa}} = \tr_B\left\{ \hat{P}_B \hat{U}_{\text{ibs}} \left[\left(\hat{U}_{\text{sq}} \hat{\rho}_{\text{in}}\hat{U}_{\text{sq}}^{\dag}\right) \otimes\hat{\rho}_{\text{vac}}\right] \hat{U}_{\text{ibs}}^{\dag} \right\} .
\end{equation}
The input state is squeezed by stimulated parametric down-conversion and, together with a vacuum state, sent through an inhomogeneous beamsplitter to separate the squeezed state into two beams, one of which is subjected to a measurement by a {\em number-resolving detector}, represented by the projection operator $\hat{P}_B$.

In terms of Wigner functionals, the stimulated parametric down-conversion process, together with the inhomogenous beamsplitter, can be performed with a {\em twin-beam Bogoliubov transformation}, which separates the signal and idler beams into two different phase spaces. To suppress the production of multiple photon pairs, the squeezing (i.e., the efficiency of the stimulated parametric down-conversion) is designed to be weak. Therefore, the squeezing parameter can be used as an expansion parameter, similar to the way we used the reflectivity as an expansion parameter in Section~\ref{aftrek}. The Bogoliubov kernels \cite{paramp} can then be expressed as
\begin{equation}
U \rightarrow \mathbf{1}+\xi^2 F ~~~~~ \text{and} ~~~~~
V \rightarrow \xi V ,
\label{swakbog}
\end{equation}
where we display the squeezing parameter $\xi$ explicitly, and $F$ represents the subleading part of $U$. The twin-beam Bogoliubov transformation is thus given by
\begin{align}
\begin{split}
\alpha \rightarrow & (\mathbf{1}+\xi^2 F)\diamond\alpha+\xi V\diamond\beta^* , \\
\beta \rightarrow & (\mathbf{1}+\xi^2 F)\diamond\beta+\xi V\diamond\alpha^* .
\end{split}
\end{align}
After applying this transformation to the combination of the Wigner functionals for the seed state and the vacuum state, we have
\begin{align}
W_{\text{ibs}}[\alpha,\beta] = & W_{\text{in}}\left[\alpha+\xi^2 F\diamond\alpha+\xi V\diamond\beta^*\right] \nonumber \\
& \times \mathcal{N}_0\exp\left[-2\left(\beta^*+\xi^2\beta^*\diamond F+\xi\alpha\diamond V^*\right) \right. \nonumber \\
& \left. \diamond\left(\beta+\xi^2 F\diamond\beta+\xi V\diamond\alpha^*\right)\right] .
\end{align}
Similar to what we had with photon subtraction, we only need to compute the second order term for single-photon addition. It reads
\begin{widetext}
\begin{align}
\tfrac{1}{2}\xi^2 \left. \partial_{\xi}^2 W_{\text{ibs}}[\alpha,\beta] \right|_{\xi=0}
= & \xi^2 \left[
\beta^*\diamond V\diamond\frac{\delta^2 W_{\text{in}}}{\delta\alpha\delta\alpha^*}\diamond V^*\diamond\beta
+\tfrac{1}{2}\beta^*\diamond V\diamond\frac{\delta^2 W_{\text{in}}}{\delta\alpha\delta\alpha}\diamond V\diamond\beta^*
+\tfrac{1}{2}\beta\diamond V^*\diamond\frac{\delta^2 W_{\text{in}}}{\delta\alpha^*\delta\alpha^*}
\diamond V^*\diamond\beta \right. \nonumber \\
& -2\left(\alpha\diamond V^*\diamond\beta+\beta^*\diamond V\diamond\alpha^*\right)
\left(\frac{\delta W_{\text{in}}}{\delta\alpha^*}\diamond V^*\diamond\beta
+\beta^*\diamond V\diamond\frac{\delta W_{\text{in}}}{\delta\alpha}\right) \nonumber \\
& - 2\left(2\beta^*\diamond F\diamond\beta+\alpha\diamond V^*\diamond V\diamond\alpha^*\right) W_{\text{in}}
+2\left(\alpha\diamond V^*\diamond\beta+\beta^*\diamond V\diamond\alpha^*\right)^2 W_{\text{in}} \nonumber \\
& \left. +\frac{\delta W_{\text{in}}}{\delta\alpha}\diamond F\diamond\alpha
+\alpha^*\diamond F\diamond\frac{\delta W_{\text{in}}}{\delta\alpha^*}\right] \mathcal{N}_0 \exp\left(-2\beta^*\diamond\beta\right) .
\end{align}

We again use Eq.~(\ref{fotnumgen2}) for the photon measurements. As in Section~\ref{aftrek}, the functional integration over $\beta$ produces a superposition of moment integrals, which can be computed with the iad of Eq.~(\ref{genmomente}). It leads to
\begin{align}
\mathcal{W}_{1\text{pa}} 
= & \tr\left\{F\diamond\frac{\delta^2 W_{\text{in}}}{\delta\alpha^*\delta\alpha}\right\}
-\frac{\delta W_{\text{in}}}{\delta\alpha}\diamond F\diamond\alpha
-\alpha^*\diamond F\diamond\frac{\delta W_{\text{in}}}{\delta\alpha^*}
-2\tr\left\{F\right\} W_{\text{in}} \nonumber \\
& -\tfrac{1}{4}(1-J)\left( \tr\left\{V^*\diamond D\diamond V\diamond\frac{\delta^2 W_{\text{in}}}{\delta\alpha\delta\alpha^*}\right\}
-4\tr\left\{D\diamond F\right\} W_{\text{in}}
+4\alpha\diamond V^*\diamond D\diamond V\diamond\alpha^* W_{\text{in}} \right. \nonumber \\
& \left. -2\alpha\diamond V^*\diamond D\diamond V\diamond\frac{\delta W_{\text{in}}}{\delta\alpha}
-2\frac{\delta W_{\text{in}}}{\delta\alpha^*}\diamond V^*\diamond D\diamond V\diamond\alpha^* \right) .
\end{align}
\end{widetext}
Here we dropped the factor of $\xi^2$ and we replaced $V\diamond V^*\rightarrow 2F$. This identity can be derived under weak squeezing conditions from Eq.~(\ref{idsqut}) and the relationships between the Bogoliubov kernels and the squeezed state kernels given by
\begin{equation}
A = U\diamond U + V\diamond V^* ~~~ \text{and} ~~~
B = U\diamond V + V\diamond U^* .
\end{equation}

Being linear in the generating parameter $J$, the generating function only allows single photon addition. Higher order expansions are required for the addition of more photons. The expression for the single-photon added state with a single-mode detector kernel reads
\begin{align}
W_{1\text{pa}} 
= & \mathcal{N} \left[ \tfrac{1}{4}M^*\diamond V\diamond \frac{\delta^2 W_{\text{in}}}{\delta\alpha\delta\alpha^*}
\diamond V^*\diamond M \right. \nonumber \\
& +\left(\alpha\diamond V^*\diamond MM^*\diamond V\diamond\alpha^*-M^*\diamond F\diamond M\right) W_{\text{in}} \nonumber \\
& -\tfrac{1}{2}\alpha\diamond V^*\diamond MM^*\diamond V\diamond\frac{\delta W_{\text{in}}}{\delta\alpha} \nonumber \\
& \left. -\tfrac{1}{2}\frac{\delta W_{\text{in}}}{\delta\alpha^*}\diamond V^*\diamond MM^*\diamond V\diamond\alpha^* \right] ,
\label{eenfotop}
\end{align}
where $\mathcal{N}$ is the normalization constant, which is required because heralded photon addition is not a trace-preserving process. With the aid of the same calculation used in Section~\ref{aftrek} to compute the normalization constant, we obtain
\begin{align}
\mathcal{N}^{-1} = & \int \alpha\diamond V^*\diamond D\diamond V\diamond\alpha^* W_{\text{in}}\ \Dcirc[\alpha]  \nonumber \\
& + \tr\left\{D\diamond F\right\} .
\end{align}
It again requires knowledge of the initial Wigner functional to evaluate the functional integral.

\subsection{Single-photon added coherent state}

Contrary to the situation with photon subtraction, photon addition applied to most states produces Wigner functionals with negative regions. We substitute the coherent state Wigner functional
\begin{equation}
W_{\text{coh}}[\alpha] = \mathcal{N}_0 \det\{T\} \exp\left[-2(\alpha^*-\xi^*)\diamond(\alpha-\xi)\right] ,
\label{kohwig}
\end{equation}
with parameter function $\xi$, into Eq.~(\ref{eenfotop}). Then we compute the normalization constant, with a generating functional for the moments, produced in the same way we did in Eq.~(\ref{genmomente}). For the coherent state, this generating functional is given by
\begin{align}
\mathcal{W}_{\mathcal{N}}[\mu^*,\nu]
= & \int \mathcal{N}_0 \exp\left(-2(\alpha^*-\xi^*)\diamond(\alpha-\xi) \right. \nonumber \\
& \left. +\alpha^*\diamond\nu+\mu^*\diamond\alpha\right)\ \Dcirc[\alpha] \nonumber \\
= & \exp\left(\xi^*\diamond\nu+\mu^*\diamond\xi+\tfrac{1}{2}\mu^*\diamond\nu\right) .
\label{gennorm}
\end{align}
leading to
\begin{equation}
\left. \frac{\delta^2 \mathcal{W}_{\mathcal{N}}}{\delta\nu(\mathbf{k})\delta\mu^*(\mathbf{k}')} \right|_{0} 
= \xi^*(\mathbf{k})\xi(\mathbf{k}')+\tfrac{1}{2}\mathbf{1}(\mathbf{k}',\mathbf{k}) .
\label{aamoment}
\end{equation}
So, the normalization constant comes out to be
\begin{equation}
\mathcal{N}^{-1} 
= M_V^*\diamond\xi\xi^*\diamond M_V + \|M_V\|^2 ,
\end{equation}
where the transformed detector mode is $M_V=V\diamond M^*$. The resulting expression for the Wigner functional of the normalized photon-added coherent state is
\begin{align}
W_{1\text{pa}}^{(\text{coh})}
= & \frac{M_V^*\diamond(2\alpha-\xi)(2\alpha^*-\xi^*)\diamond M_V-\|M_V\|^2}{M_V^*\diamond\xi\xi^*\diamond M_V+\|M_V\|^2} \nonumber \\
& \times W_{\text{coh}} ,
\label{koheenfotop}
\end{align}
This Wigner functional is negative at $\alpha=\tfrac{1}{2}\xi$. However, for $\|\xi\|\gg 1$, the negative amplitude of the Wigner functional at $\alpha=\tfrac{1}{2}\xi$ is severely suppressed by the Gaussian shape of the coherent state Wigner functional \cite{zav1}.

\begin{figure}[ht]
\centerline{\includegraphics{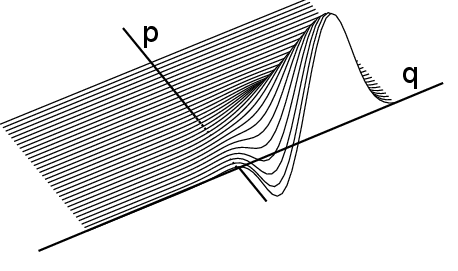}}
\caption{Wigner function of photon-added coherent state.}
\label{kohfotop}
\end{figure}

The polynomial factor in Eq.~(\ref{koheenfotop}) exists in a subspace of the functional phase space associated with a single mode, namely $M_V$. Therefore, only the parts of the field variable $\alpha$ and the parameter function $\xi$ that are proportional to $M_V$ play a role in the shape of the Wigner functional. We can substitute $M_V\rightarrow m_0 G$, $\alpha\rightarrow\alpha_0 G$ and $\xi\rightarrow\xi_0 G$, where $G$ is a normalized mode, $\alpha_0$ is a complex variable and $\xi_0$ and $m_0$ are constants. After tracing over the part of the functional phase space that is orthogonal to $G$, we end up with a two-dimensional Wigner function (not a functional anymore), given by
\begin{align}
W_{1\text{pa}}^{(\text{coh})}(\alpha_0) = & 2\frac{|2\alpha_0-\xi_0|^2-1}{|\xi_0|^2+1} \exp\left(-2 |\alpha_0-\xi_0|^2\right) .
\end{align}
The magnitude of the transformed mode $|m_0|=\|M_V\|$ cancels out everywhere. If we assume that the coherent state is weak so that $|\xi_0|\approx 1$, we can plot the two-dimensional Wigner function of the single-photon added coherent state. It is shown in Fig.~\ref{kohfotop}, as a function of $q$ and positive $p$ to reveal the negative region lying between the origin and maximum of the Wigner function.

\subsection{Single-photon added thermal state}

We also consider a thermal state as initial state for the photon-addition process \cite{zav3}. Its Wigner functional is
\begin{equation}
W_{\text{th}}[\alpha] = \mathcal{N}_0 \det\{T\} \exp(-2\alpha^*\diamond T\diamond\alpha) ,
\label{termwig}
\end{equation}
where $T$ is the thermal state kernel. Substituted into Eq.~(\ref{eenfotop}), with a single-mode detector kernel, it gives
\begin{align}
W_{1\text{pa}}^{(\text{th})}
= & \mathcal{N} \left[M_V^*\diamond(T+\mathbf{1})\diamond\alpha \alpha^*\diamond(T+\mathbf{1})\diamond M_V \right. \nonumber \\
& \left. -\tfrac{1}{2} M_V^*\diamond(T+\mathbf{1})\diamond M_V\right] W_{\text{th}} .
\label{theenfotop}
\end{align}
Following the same steps as before, we produce a generating functional for the moments, given by
\begin{align}
\mathcal{W}_{\mathcal{N}}[\mu^*,\nu]
= & \int \det\{T\} \exp\left(-2\alpha^*\diamond T\diamond\alpha \right. \nonumber \\
& \left. +\alpha^*\diamond\nu+\mu^*\diamond\alpha\right)\ \Dcirc[\alpha] \nonumber \\
= & \exp\left(\tfrac{1}{2}\mu^*\diamond T^{-1}\diamond\nu\right) ,
\end{align}
so that
\begin{equation}
\left. \frac{\delta^2 \mathcal{W}_{\mathcal{N}}}{\delta\nu(\mathbf{k})\delta\mu^*(\mathbf{k}')} \right|_{0} 
= \tfrac{1}{2}T^{-1}(\mathbf{k}',\mathbf{k}) .
\end{equation}
The normalization constant is then given by
\begin{equation}
\mathcal{N}^{-1} 
= \tfrac{1}{2} M_V^*\diamond(T^{-1}+\mathbf{1})\diamond M_V .
\end{equation}
The normalized photon-added thermal state is
\begin{align}
W_{1\text{pa}}^{(\text{th})}
= & \frac{M_V^*\diamond(T+\mathbf{1})\diamond\alpha \alpha^*\diamond(T+\mathbf{1})\diamond M_V-\Lambda_0}{\Lambda_1} \nonumber \\
& \times W_{\text{th}} .
\end{align}
where
\begin{align}
\begin{split}
\Lambda_0 = & \tfrac{1}{2} M_V^*\diamond(T+\mathbf{1})\diamond M_V , \\
\Lambda_1 = & \tfrac{1}{2} M_V^*\diamond(T^{-1}+\mathbf{1})\diamond M_V .
\end{split}
\end{align}

\begin{figure}[ht]
\centerline{\includegraphics{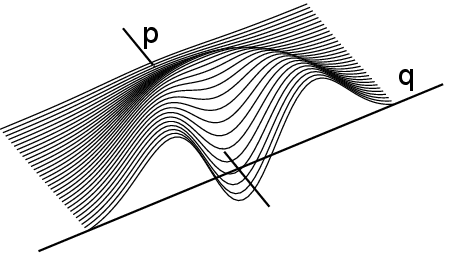}}
\caption{Wigner function of photon-added thermal state.}
\label{thfotop}
\end{figure}

For simplicity, we consider a single-mode thermal state, which is often encountered in experiments where the thermal source is represented by a beam of thermal light with a specific spatiotemporal mode. The kernel of such a single-mode thermal state can be represented by
\begin{equation}
T = \left(\mathbf{1}+\tau \Theta\Theta^*\right)^{-1}
= \mathbf{1}-\frac{\tau}{1+\tau} \Theta\Theta^* ,
\end{equation}
where $\tau$ is the average number of photons in the thermal state and $\Theta$ is a normalized mode. The determinant becomes $\det\{T\}=(1+\tau)^{-1}$. Assuming that the transformed detector mode is proportional to the mode of the thermal state, we replace $M_V\rightarrow m_0 \Theta$. With the part orthogonal to $\Theta$ traced out, the photon-added thermal state's Wigner function becomes
\begin{align}
W_{1\text{pa}}^{(\text{th})}(\alpha_0)
= & \frac{2}{(1+\tau)^2}\left(2\frac{2+\tau}{1+\tau}|\alpha_0|^2-1\right)\nonumber \\
& \times \exp\left(-\frac{2|\alpha_0|^2}{1+\tau}\right) ,
\end{align}
where $\alpha_0=\Theta^*\diamond\alpha$. In Fig.~\ref{thfotop}, the Wigner function of the photon-added thermal state is shown for $\tau=5$, plotted as a function of $q$ and positive $p$. It reveals the negative region at the origin.

\section{\label{concl}Conclusions}

The effect of the spatiotemporal degrees of freedom in the experimental implementations of heralded quantum state preparation is investigated. The Wigner functional formalism is used to derive general expressions for the states produced by heralded photon subtraction and heralded photon addition. While the initial states are parameterized in terms of kernels and parameter functions, the heralding measurement is represented by a detector kernel, which can be parameterized in terms of a single spatiotemporal detector mode.

As examples, the heralded preparation of photon-subtracted squeezed vacuum states, photon-added coherent states, and photon-added thermal states are considered. In all these cases, the single-mode measurements reduce the dimensionality of the Wigner functional of the heralded state. The Wigner functional of the photon-subtracted squeezed vacuum states is represented within a four-dimensional subspace of the functional phase space spanned by two complex-valued transformed detector modes, because the initial squeezed vacuum states is parameterized by two kernels. Both the examples of photon-added states are rendered in two-dimensional subspaces of the functional phase space, represented as a complex plane associated with the complex-valued transformed detector mode. The reason is that the heralded photon addition process is based on a measurement of the difference-frequency component of the Bogoliubov transformed input state. It thus involves only one Bogoliubov kernel. The two-dimensional Wigner functions of these photon-added states can therefore be shown in plots as provided.

Although the domains of these heralded states are significantly reduced by the single-mode heralding measurements, there are still enough variability in the properties of the detector modes and the initial states to provide much diversity for the properties for the heralded states. With the aid of this analysis, one can investigate these properties for optimal design of the quantum information systems involved.

One of the important aspects resulting from this analysis is that the finite-dimensional domain on which the heralded state's Wigner function is represented, is defined by transformed detector modes. When such a Wigner function is sent into a subsequent system for further processing (for example, being measured with the aid of homodyne tomography), it is important that the spatiotemproal degrees of freedom of the modes used in such a subsequent process (such as the local oscillator in the homodyne tomography system) match the {\em transformed} detector mode of the heralding detector, and not the original heralding detector mode. Any mismatch between these modes and the transformed detector mode would lead to diminished or distorted results \cite{homod}.



\end{document}